\documentclass[onecollarge,natbib]{svjour2}
\bibpunct{[}{]}{;}{n}{}{,} 
\smartqed  
\usepackage{slashed}
\usepackage{graphicx}
\usepackage{placeins}
\usepackage{color}
\usepackage{latexsym}
\usepackage{amsmath}
\newcommand{\be}{\begin{equation}}
\newcommand{\ee}{\end{equation}}
\newcommand{\bea}{\begin{eqnarray}}
\newcommand{\eea}{\end{eqnarray}}

\newcommand{\bfk}{\mbox{\boldmath $k$}}
\newcommand{\bfq}{\mbox{\boldmath $q$}}

\newcommand{\pup}{p^\uparrow}

\newcommand{\kt}{k_\perp}

\def\lsim{\mathrel{\rlap{\lower4pt\hbox{\hskip1pt$\sim$}}\raise1pt\hbox{$<$}}}
\def\gsim{\mathrel{\rlap{\lower4pt\hbox{\hskip1pt$\sim$}}\raise1pt\hbox{$>$}}}

\journalname{Few-Body Systems}
\begin{document}

\title{Single Spin Asymmetry in Charmonium Production
}


\author{Rohini M. Godbole     \and Abhiram Kaushik  \and 
        Anuradha Misra  \and  Vaibhav Rawoot 
}

\institute{Anuradha Misra  \at
Department of Physics, University of Mumbai, 
SantaCruz(East), Mumbai-400098, India. \\
\email misra@physics.mu.ac.in
\\           
           \and
           Rohini Godbole \and Abhiram Kaushik    \at
           Indian Institute of Science, Bangalore, India-560012\\
\email rohini@cts.iisc.ernet.in,  abhiramb@cts.iisc.ernet.in 
\\
\and
Vaibhav Rawoot \at
Institute of Mathematical Sciences, Chennai, India-600113 \\
\email vaibhavrawoot@gmail.com
}
%
        
\date{Received: date / Accepted: date}

\maketitle

\begin{abstract}
We present estimates of Single Spin Asymmetry (SSA) in the electroproduction of $J/\psi$ taking into account the  transverse momentum dependent (TMD) evolution 
of the gluon Sivers function and using Color Evaporation Model of charmonium production.  We estimate SSA for JLab, HERMES, COMPASS and eRHIC energies  using recent parameters for the quark Sivers functions which are  fitted using an 
evolution kernel in which the perturbative part is  resummed up to  next-to-leading logarithms (NLL) accuracy. We find that these SSAs are  much smaller as compared to our first estimates obtained using DGLAP evolution but are comparable to our estimates obtained using TMD evolution where we had used approximate analytical solution of the TMD evolution equation for the purpose.
\end{abstract}

\section{Introduction}
\label{intro}
Transverse Single Spin Asymmetries (SSA's) arise  in the scattering of a transversely polarized nucleon 
off an unpolarized nucleon (or virtual photon) target 
when the final observed hadrons have asymmetric distribution in the transverse plane perpendicular to the beam direction depending on the polarization vector of the scattering nucleon. One of the two major theoretical approaches to explain these asymmetries is the Transverse Momentum Dependent (TMD) approach\cite{Collins:2011book,Aybat:2011zv} which is based on a pQCD factorization scheme which  includes the spin and TMD effects in the collinear factorization scheme. An important Transverse Momentum Dependent Distribution (TMD) is the Sivers function which 
is related to  the density of unpolarized 
partons in a transversely polarized nucleon.

The number density of partons inside proton with transverse polarization
${\bf S}$, three momentum ${\bf p}$ and intrinsic transverse momentum ${\bf k_\perp}$ of partons,
is expressed in terms of the Sivers function, $\Delta^{N} 
f_{a/p} (x,k_{\perp})$, as
\begin{equation}
\hat f_{a/p^\uparrow}(x, {\bf k}_{\perp}) = \hat f_{a/p}(x, k_{\perp})
+ \frac{1}{2} \Delta^N f_{a/p^\uparrow}(x, k_{\perp}) 
{\bf S} \cdot (\hat{\bf p} \times \hat{\bf k}_{\perp})
\end{equation}

Heavy quark and quarkonium systems are natural probes to study gluon Sivers function as the production and the differential distributions of the produced charmonium are directly dependent on the intrinsic transverse momentum  distribtuion of the gluon especially at low transverse momentum\cite{Yuan:2008vn}.
 We have proposed study of SSA in $J/\psi$ production as a possible probe of gluon Sivers function and have estimated SSA in photo production (i.e. low virtuality electro production) of $J/\psi$ in scattering of electrons off transversely polarized protons,   
using the color evaporation model (CEM) of charmonium production\cite{Godbole:2012bx,Godbole:2013bca}. 
Here,  we present some  preliminary results containing improved estimates of asymmetry 
 taking into account the TMD evolution of the Sivers function up to next-to-leading logarithm (NLL) order. A more detailed analysis can be found in Ref. \cite{Godbole:2014tha}.
 
 \section{Transverse Single Spin Asymmetry in $e+\pup\rightarrow J/\psi +X$}
 In the process under consideration, at LO, there is contribution only from a single
partonic subprocess $\gamma g\rightarrow c\bar{c}$ and hence it provides a clean probe of gluon Sivers function.
The CEM expression for electroproduction of $J/\psi$ can be generalized by 
taking into account the transverse momentum dependence
of the William Weizsaker (WW) function and gluon distribution function and can be written as
\begin{equation}
\sigma^{e+p^\uparrow\rightarrow e+J/\psi + X}=
\int_{4m_c^2}^{4m_D^2} dM_{c\bar c}^2\> dx_\gamma\> dx_g\> d^2\bfk_{\perp\gamma}d^2\bfk_{\perp g}\>
f_{g/p^{\uparrow}}(x_{g},\bfk_{\perp g}) \nonumber \\
f_{\gamma/e}(x_{\gamma},\bfk_{\perp\gamma})\>
\frac{d\hat{\sigma}^{\gamma g\rightarrow c\bar{c}}}{dM_{c\bar c}^2} \nonumber
\label{dxec-ep}
\end{equation}
where $f_{\gamma/e}(y,E)$ is the distribution function of the photon in the electron given by 
William Weizsaker approximation \cite{Kniehl}.
We assume $\kt$ dependence of pdf's and WW function to be factorized in gaussian form \cite{Anselmino:2008sga,Godbole:2012bx}.
\be
f(x,k_{\bot})=f(x)\frac{1}{\pi\langle k^{2}_{\bot}\rangle} 
e^{-k^{2}_{\bot}/\langle{k^{2}_{\bot}\rangle}} \quad\quad\quad\quad 
\langle k^{2}_{\bot}\rangle=0.25 GeV^2 \nonumber
\label{gauss}
\ee 
We use  the following  model for gluon Sivers function proposed by Anselmino {\it et al.}\cite{Anselmino:2008sga}
\be
 \Delta^Nf_{g/\pup}(x,\kt) = 2 {\mathcal N}_g(x)\sqrt{2e} \frac{{\kt}}{M_{1}} e^{-{{\kt}^2}/{M_{1}^2}} f_{g/p}(x)
\frac{e^{-k^{2}_{\bot}/\langle{k^{2}_{\bot}\rangle}}}{\pi\langle k^{2}_{\bot}\rangle} 
\cos{\phi_{k_\perp}} \nonumber
\label{dnf}
\ee
${\mathcal N}_g(x)$ is the $x$-dependent normalization for which we have used
${\mathcal N}_g(x)={\mathcal N}_d(x) \;$. Since there is no information on gluon Sivers function, one parameterizes the gluon Sivers function in terms of quark Sivers function.
 The $x$-dependent normalization for  quarks is \cite{Anselmino:2008sga}
\be 
{\mathcal N}_f(x) = N_f x^{a_f} (1-x)^{b_f} \frac{(a_f + b_f)^{(a_f +
b_f)}}{{a_f}^{a_f} {b_f}^{b_f}} \nonumber
\label{siversx} 
\ee
where $a_f, b_f$ and $N_f$ are best fit parameters.
The weighted Sivers asymmetry integrated over the azimuthal angle of 
$J/\psi$\cite{Collins:2005rq} is given by

\be
A_N=\frac{\int d\phi_{q}\int_{4m^2_c}^{4m^2_D}[dM^{2}] \int d^2\bfk_{\perp g}
{\Delta^{N}f_{g/\pup}(x_{g},\bfk_{\perp g})}
f_{\gamma/e}(x_{\gamma},\bfq_T-\bfk_{\perp g})
\hat\sigma_{0}sin(\phi_{q}-\phi_S)}
{2\int d\phi_{q}\int_{4m^2_c}^{4m^2_D}[dM^{2}]\int d^{2}\bfk_{\perp g} f_{g/P}(x_g,\bfk_{\perp g})f_{\gamma/e}(x_{\gamma},\bfq_T-\bfk_{\perp g})\hat{\sigma}_0} 
\label{an2}
\ee
where $\phi_q$ and $\phi_S$ are the azimuthal angles of the $J/\psi$ and proton spin respectively.  The weight factor is $\sin(\phi_{q}-\phi_S)$ and $x_{g,\gamma} = \frac{M}{\sqrt s} \, e^{\pm y} $.

\section{QCD evolution of TMDs}
Early  phenomenological fits of Sivers function  were performed either  neglecting QCD 
evolution or applying DGLAP evolution only to the collinear part of TMD parametrization.
Recently a TMD factorization formalism has been derived and implemented by Collins {\it et al.} \cite{Collins:2011book}.
TMD evolution describes how the form of distribution changes and also how the 
width changes in momentum space.
 A strategy to extract Sivers function from SIDIS data taking into account the TMD $Q^2$ evolution was
 proposed by Anselmino {\it et al.}\cite{Anselmino:2012aa}.
 In our earlier work, we  estimated SSA in electroproduction of $J/\psi$ using this strategy
\cite{Godbole:2013bca}.
The energy evolution of a general transverse momentum dependent distribution(TMD) $F(x,k_\perp,Q)$ is more naturally described in b-space.
The TMDPDF in $b$-space evolves according to 
\be
 F(x,b,Q_f)=F(x,b,Q_i)R_{pert}(Q_f,Q_i,b_*)R_{NP}(Q_f,Q_i,b)
\label{evolution}
\ee 
where $R_{pert}$ is the perturbative part of the evolution kernel, $R_{NP}$ is the  non-perturbative part and $b_*=b/\sqrt{1+(b/b_\text{max})^2}$. The perturbative part is given by
\be
R_{pert}(Q_f,Q_i,b)=\exp\left\{-\int_{Q_i}^{Q_f}\frac{d\mu}{\mu}\left(A\ln\frac{Q_f^2}{\mu^2}+B\right)\right\}\left(\frac{Q_f^2}{Q_i^2}\right)^{-D(b;Q_i)}
\ee
where  $\frac{dD}{d\ln\mu}=\Gamma_\text{cusp}$.  
The non-perturbative exponential part contains a Q-dependent factor universal to all TMD's ($g_2$), and a factor which gives the gaussian width in $b$-space of the particular TMD ($g_1$). 
\be
R_{NP}=\exp\left\{-b^2\left(g_1^\text{TMD}+\frac{g_2}{2}\ln\frac{Q_f}{Q_i}\right)\right\}  
\nonumber
\ee
The $Q^2$-dependent  TMD's in momentum space are obtained by Fourier transforming $F(x,b,Q_f)$.
The perturbative evolution kernel $ R(Q, Q_0, b)$, which drives the $Q^2$-evolution of TMD's, 
becomes independent of $b$ in the limit $b \rightarrow \infty$ i.e. $R(Q, Q_0, b )\rightarrow R(Q, Q_0)$. 
 The $b$  integration can then be performed analytically and $Q^2$ dependent PDF's can be obtained. 
In Ref. \cite{Godbole:2013bca}, we had used this  analytical solution of approximated TMD evolution equations
given by Anselmino {\it et al.} \cite{Anselmino:2012aa} to estimate the asymmetry. 

Here, we present our improved estimates based on exact solution of evolution equations. 
Echevarria {\it et al.}\cite{Echevarria:2014xaa} have recently 
considered solution of TMD evolution equations up to NLL accuracy and have performed a global fit of all 
the experimental data on the Sivers asymmetry in SIDIS using this formalism. 
Since the  derivative of $b$-space  Sivers function satisfies the same evolution equation 
as the unpolarized PDF\cite{Aybat:2011ge}, its evolution is given by 
\begin{align}
f'^{\perp g}_{1T}(x,b;Q_f)=&\frac{M_p b}{2}T_{g,F}(x,x,Q_i)\exp\left\{-\int_{Q_i}^{Q_f}\frac{d\mu}{\mu}\left(A\ln\frac{Q^2}{\mu^2}+B\right)\right\}\left(\frac{Q_f^2}{Q_i^2}\right)^{-D(b^*;Q_i)} \nonumber\\ 
\times&\exp \left\{-b^2\left(g_1^\text{sivers}+\frac{g_2}{2}\ln\frac{Q_f}{Q_i}\right)\right\}
\end{align}
Here, $T_{q,F}(x,x,Q)$ is  the twist three  Qiu-Sterman quark gluon correlation  function which is related  to the first $k_T$ 
moment of quark Sivers function\cite{Kang:2011mr} and can be expressed in terms of the unpolarized collinear PDFs~\cite{Kouvaris:2006zy, Echevarria:2014xaa}.

\begin{equation}
T_{q,F}(x,x,Q)=\mathcal{N}_q(x)f_{q/P}(x,Q)
\end{equation}

The expansion coefficients with the appropriate gluon anomalous dimensions at
NLL are 

\begin{equation}
A^{(1)}=C_A \nonumber;
\\
A^{(2)}=\frac{1}{2}C_F\left(C_A\left(\frac{67}{18}-\frac{\pi^2}{6}\right)-\frac{5}{9}C_AN_f\right) \nonumber;
\\
B^{(1)}=-\frac{1}{2}\left(\frac{11}{3}C_A-\frac{2}{3}N_f \right) \nonumber \\;
D^{(1)}=\frac{C_A}{2}\ln\frac{Q_i^2b^2}{c^2}  
\end{equation}

Choosing the initial scale $Q_i=c/b$, the $D$ term vanishes at NLL. Taking Fourier transform of Eq. (6), one gets 
$f^{\perp g}_{1T}(x,k_\perp;Q_f)$ which is related to Sivers function through
\begin{equation}
\Delta^{N}f_{g/p^{\uparrow}}(x_{g},\bfk_{\perp g},Q)=-2\frac{k_{\perp g}}{M_p}f^{\perp g}_{1T}(x_g,k_{\perp g};Q)\cos\phi_{k_\perp}
\end{equation}

\begin{table}[t]
\begin{center}
\begin{tabular}{ | l | l | l | p{5cm} |}
 \hline
 TMD-e1 & TMD-a & TMD-e2 \\ 
 \hline 
$N_u=0.77, N_d=-1.00$&$N_u=0.75, N_d=-1.00$&$N_u=0.106,N_d=-0.163$\\
$a_u=0.68,a_d=1.11$&$a_u=0.82,a_d=1.36$&$a_u=1.051,a_d=1.552$\\
$b_u=b_d=3.1, $&$b_u=b_d=4.0, $&$b_u=b_d=4.857$, \\
$M_1^2=0.40\text{GeV}^2$&$M_1^2=0.34\text{GeV}^2$&$\langle k^2_{s\perp}\rangle = 0.282 \text{ GeV}^2$\\
$\langle k^2_{\perp}\rangle = 0.25\text{GeV}^2$&$\langle k^2_{\perp}\rangle = 0.25\text{GeV}^2$&$\langle k^2_{\perp}\rangle = 0.38\text{GeV}^2$\\
$b_{max} = 0.5 GeV^{-1}$&$b_{max} = 0.5 GeV^{-1}$&$b_{max} = 1.5 GeV^{-1}$\\
$g_2 = 0.68 \text{ GeV}^2$&$g_2 = 0.68 \text{ GeV}^2$&$g_2 = 0.16 \text{ GeV}^2$\\
 \hline
\end{tabular}
\caption{Parameter set for the Sivers function.}
 \label{parameter:set}
\end{center}
\end{table}

\begin{figure}[h]
\begin{center}
\includegraphics[width=0.32\linewidth,angle=0]{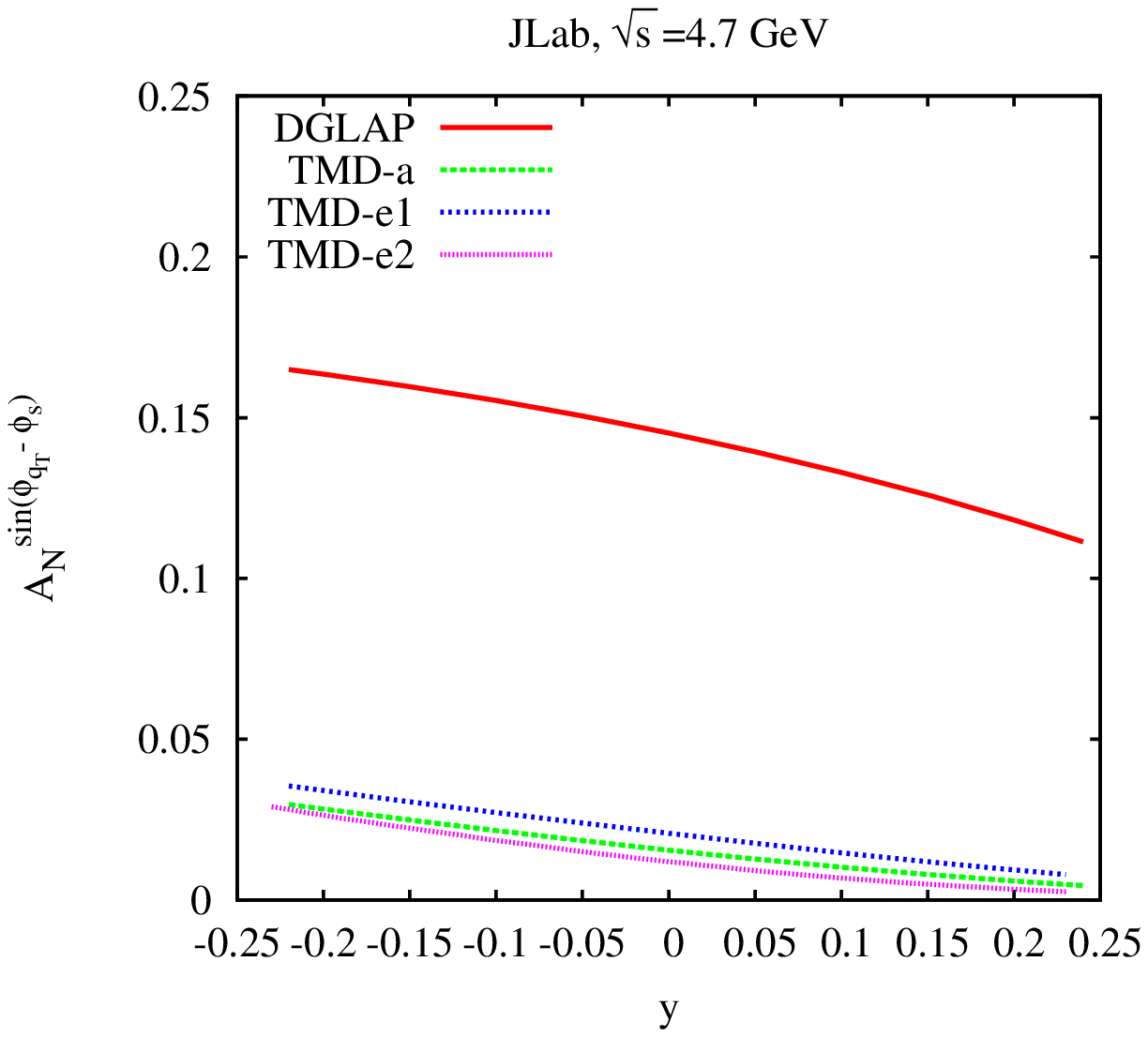}
\includegraphics[width=0.32\linewidth,angle=0]{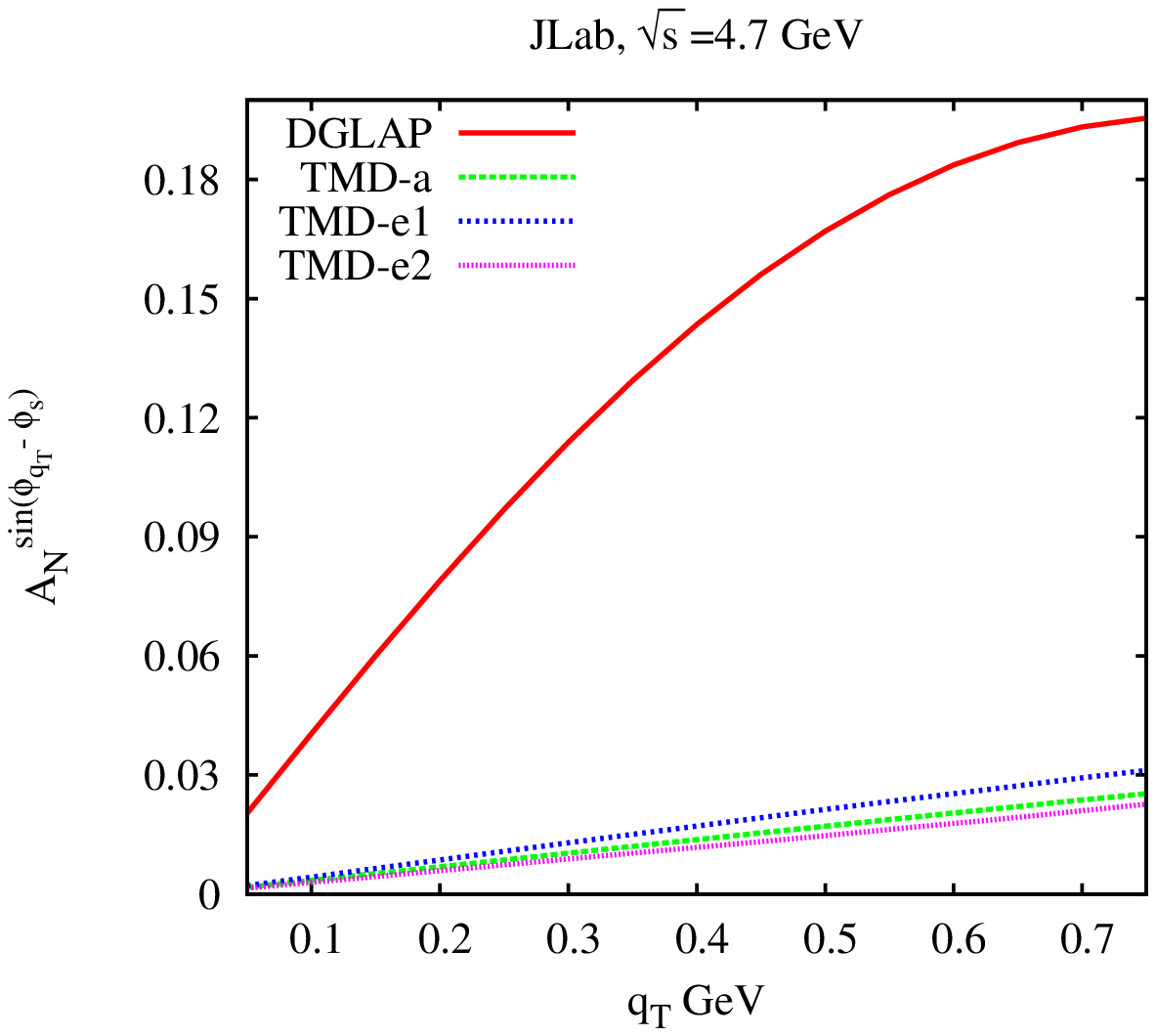}
\caption{The Sivers asymmetry $A_N^{\sin({\phi}_{q_T}-\phi_S)}$
 for $e+p^\uparrow \rightarrow  e+J/\psi +X $
at JLab energy ($\sqrt{s} = 4.7$ GeV), as a function of $y$ (left panel) and $q_T$ (right panel). The integration ranges are $(0 \leq q_T \leq 1)$~GeV and $(-0.25 \leq y \leq 0.25)$\cite{Godbole:2014tha}.} 
\label{jlab_a}
\end{center}
\end{figure}
\begin{figure}[h]
\begin{center}
\includegraphics[width=0.32\linewidth,angle=0]{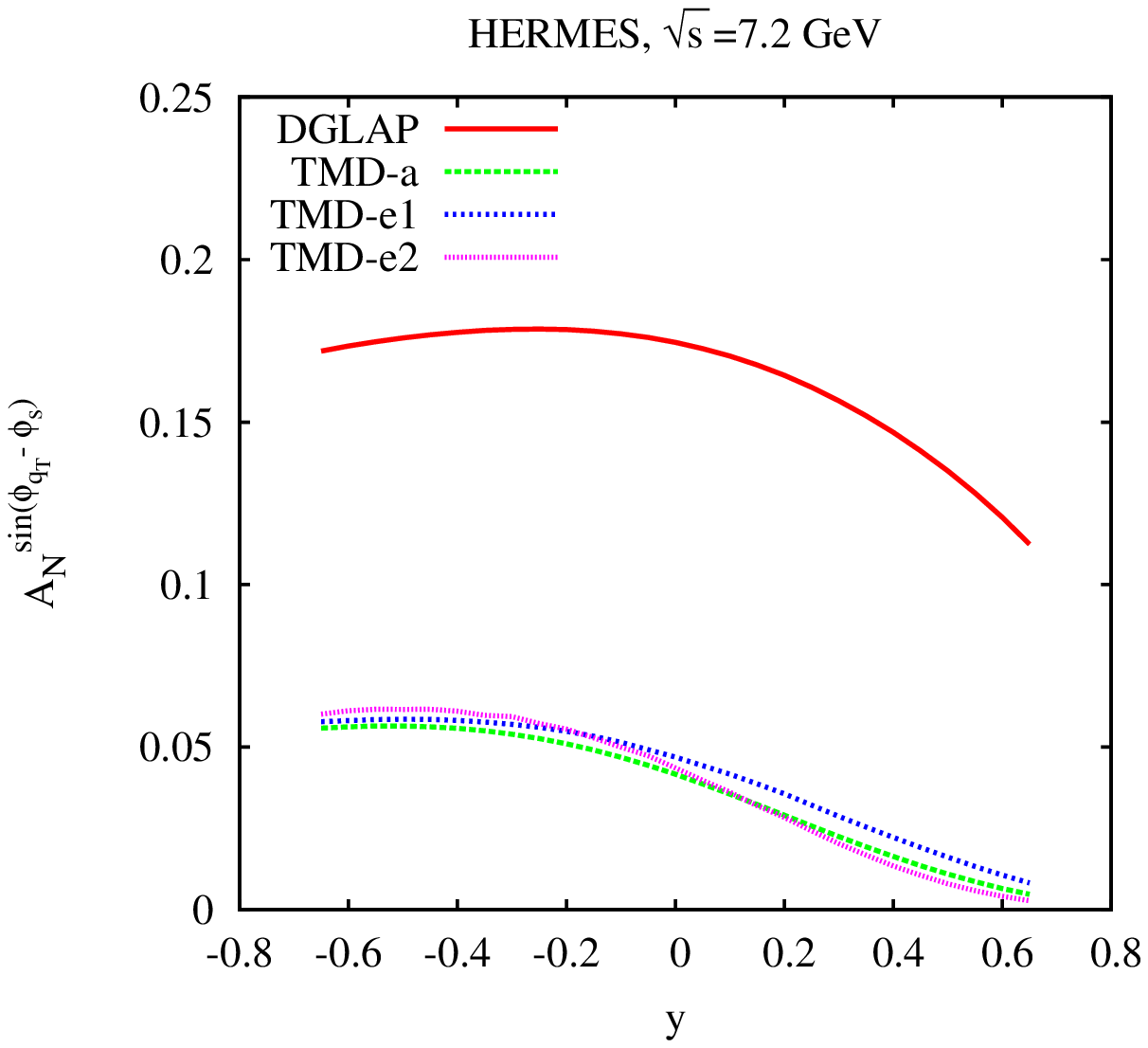}
\includegraphics[width=0.32\linewidth,angle=0]{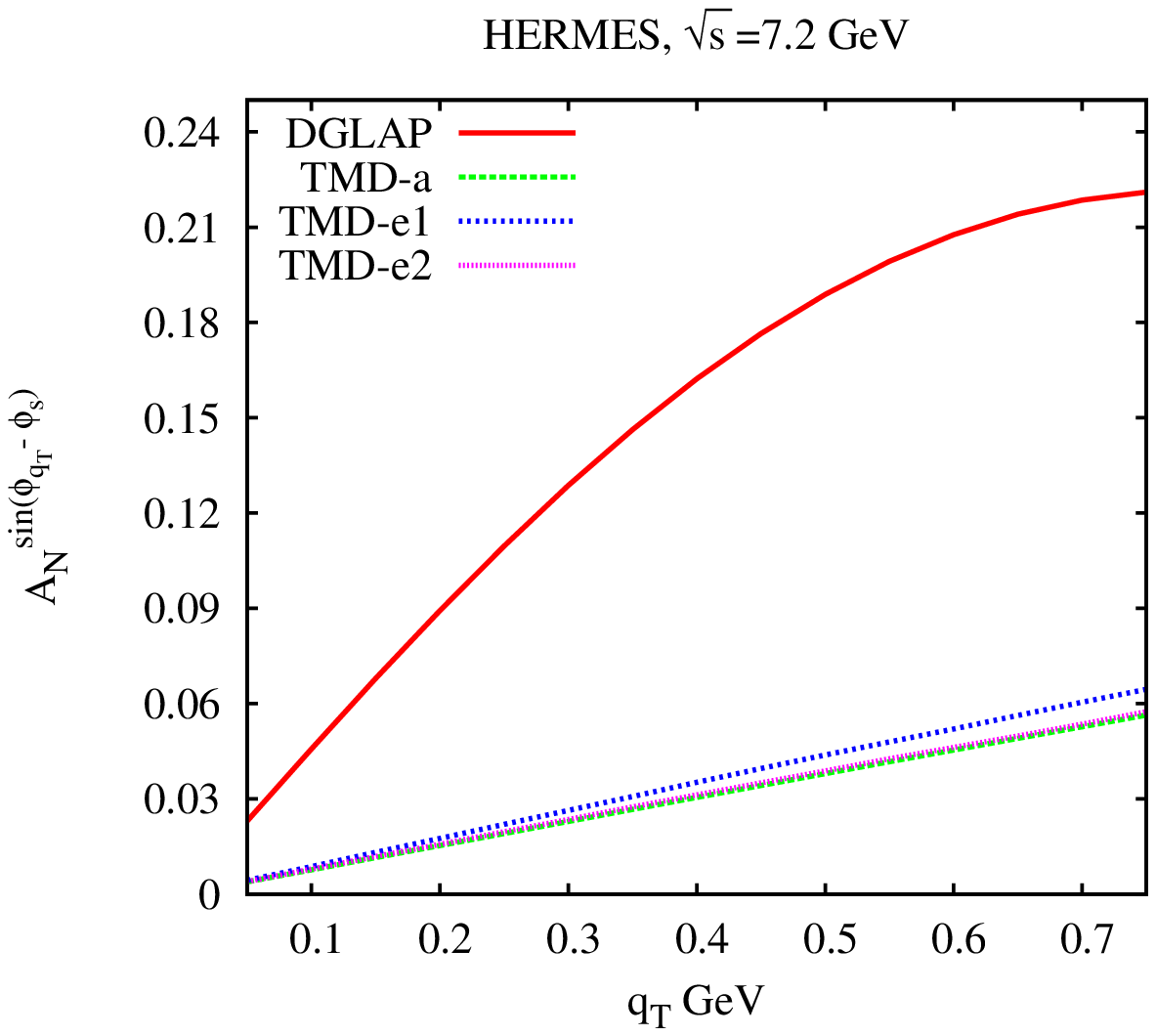}
\caption{ HERMES energy ($\sqrt{s} = 7.2$ GeV), Asymmetry as a function of $y$ (left panel) and $q_T$ (right panel). The integration ranges are $(0 \leq q_T \leq 1)$~GeV and $(-0.6 \leq y \leq 0.6)$\cite{Godbole:2014tha}. }
\label{hermes_a}
\end{center}
\end{figure}
\begin{figure}[h]
\begin{center}
\includegraphics[width=0.32\linewidth,angle=0]{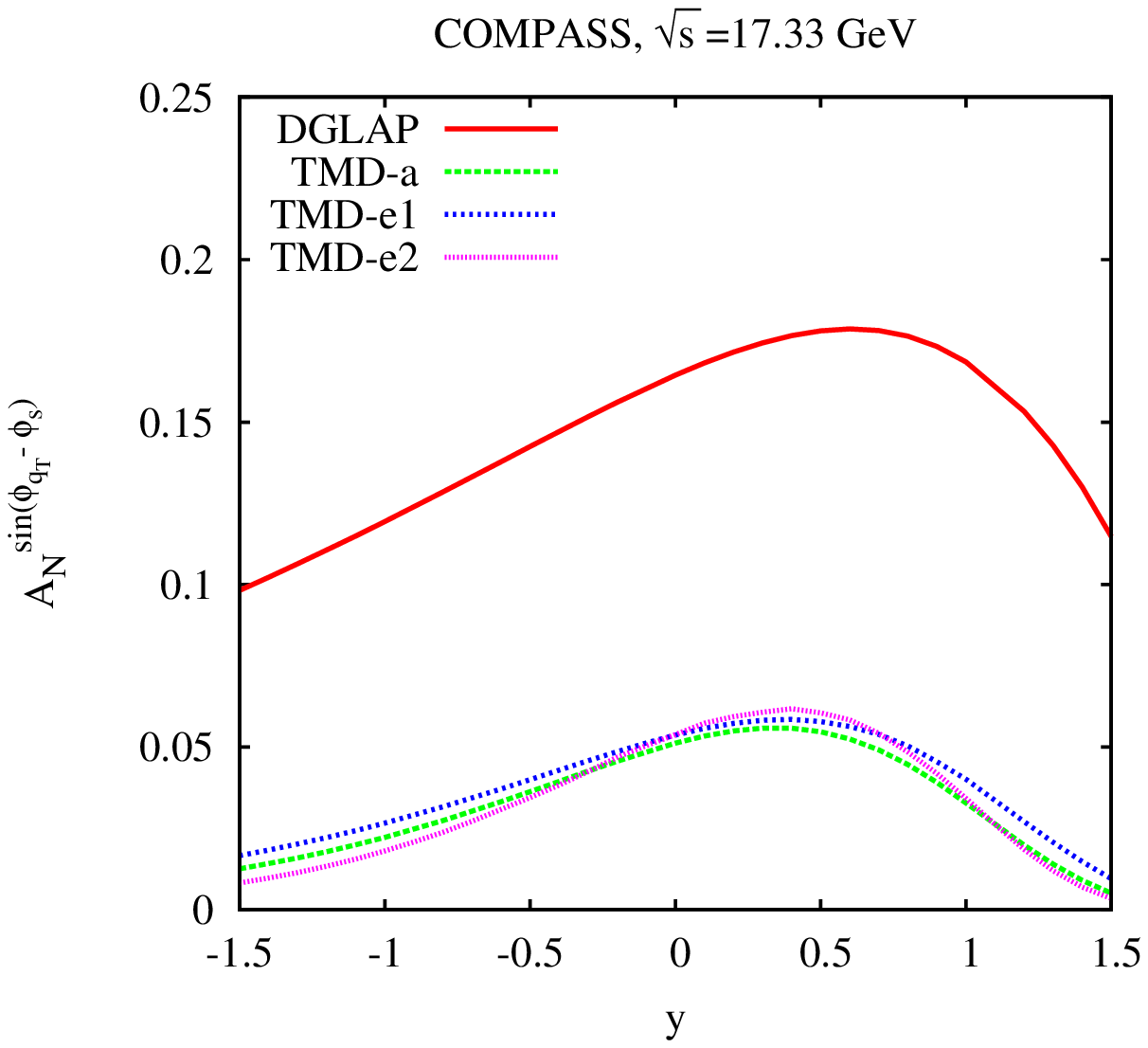}
\includegraphics[width=0.32\linewidth,angle=0]{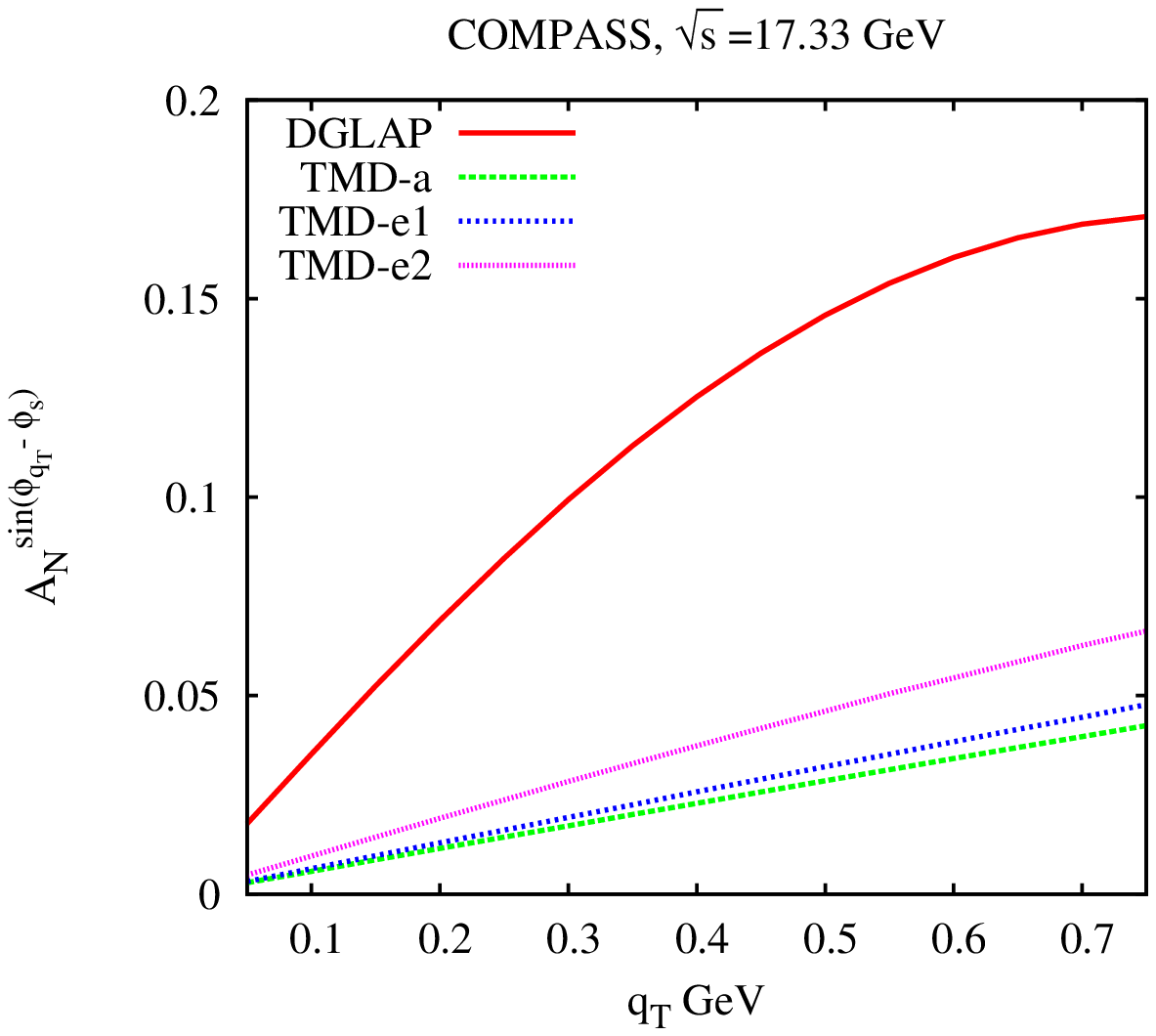}
\caption{COMPASS energy ($\sqrt{s} = 17.33$ GeV), Asymmetry as a function of $y$ (left panel) and $q_T$ (right panel). The integration ranges are $(0 \leq q_T \leq 1)$ GeV and $(-1.5 \leq y \leq 1.5)$\cite{Godbole:2014tha}.}
\label{compass_a}
\end{center}
\end{figure}
\begin{figure}[h]
\begin{center}
\includegraphics[width=0.32\linewidth,angle=0]{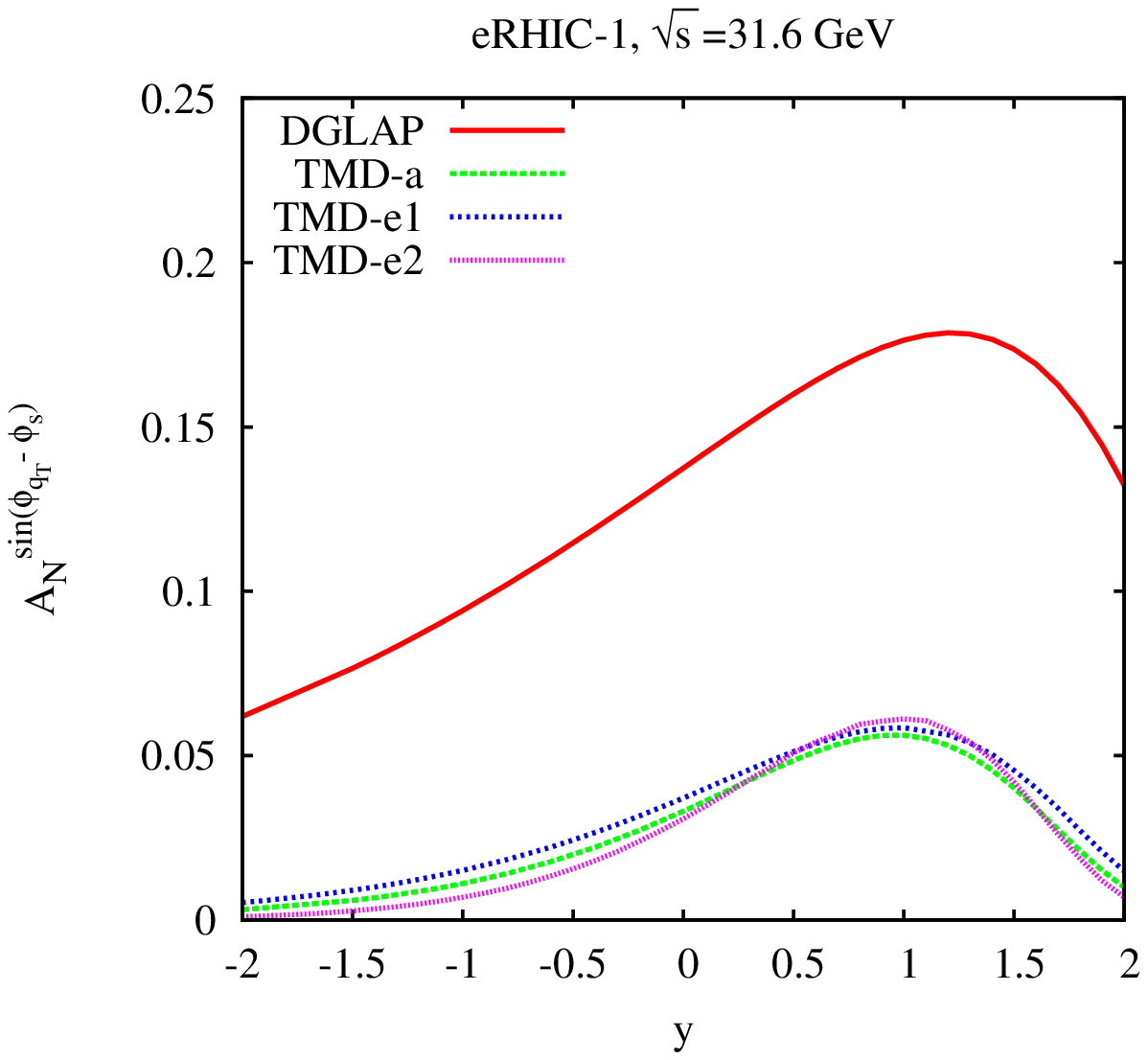}
\includegraphics[width=0.32\linewidth,angle=0]{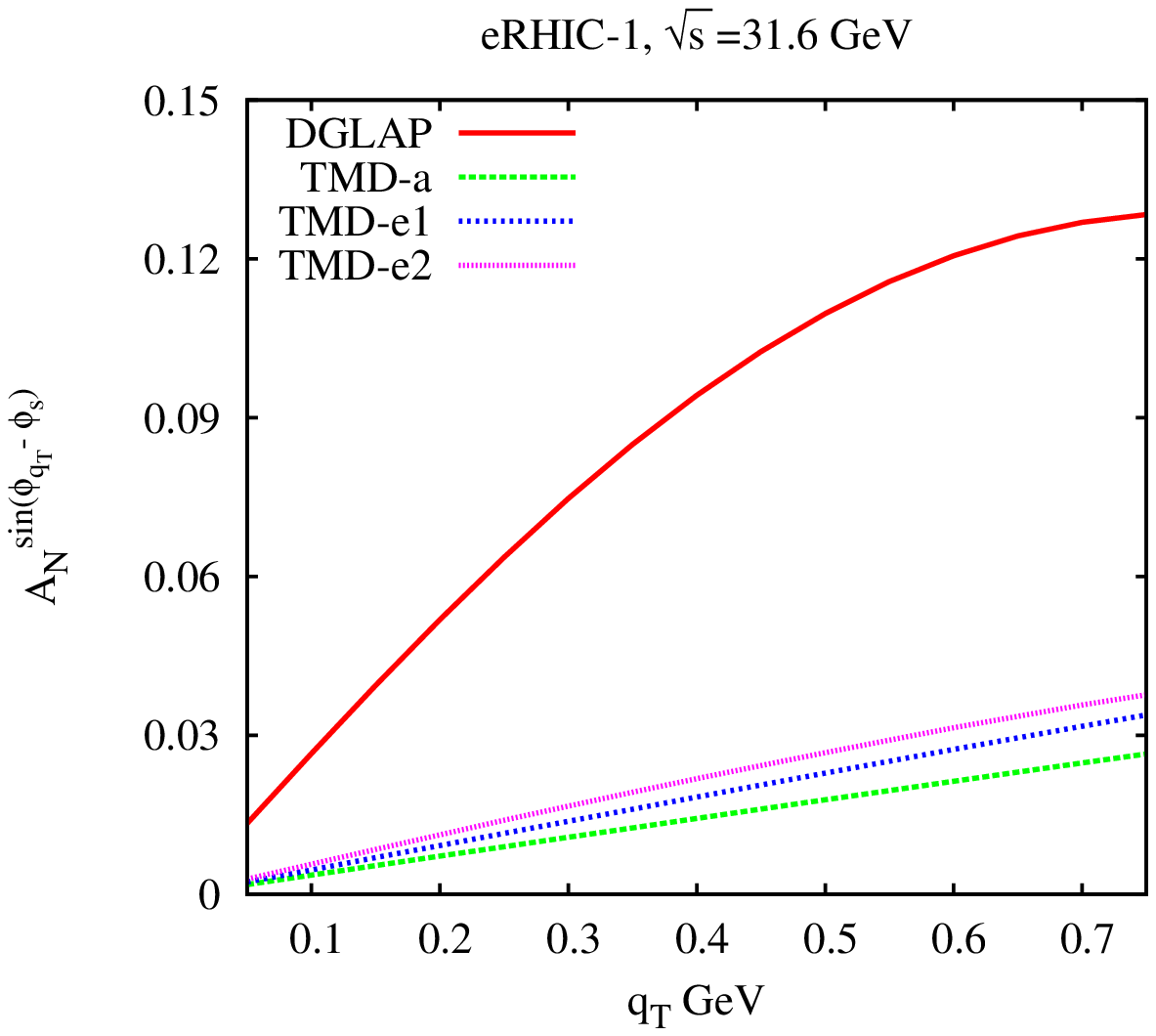}
\caption{eRHIC energy ($\sqrt{s} = 31.6$ GeV), Asymmetry as a function of $y$ (left panel) and $q_T$ (right panel). The integration ranges are $(0 \leq q_T \leq 1)$ GeV and $(-2.1 \leq y \leq 2.1)$\cite{Godbole:2014tha}. }
\label{erhic1_b}
\end{center}
\end{figure}
\begin{figure}[h]
\begin{center}
\includegraphics[width=0.32\linewidth,angle=0]{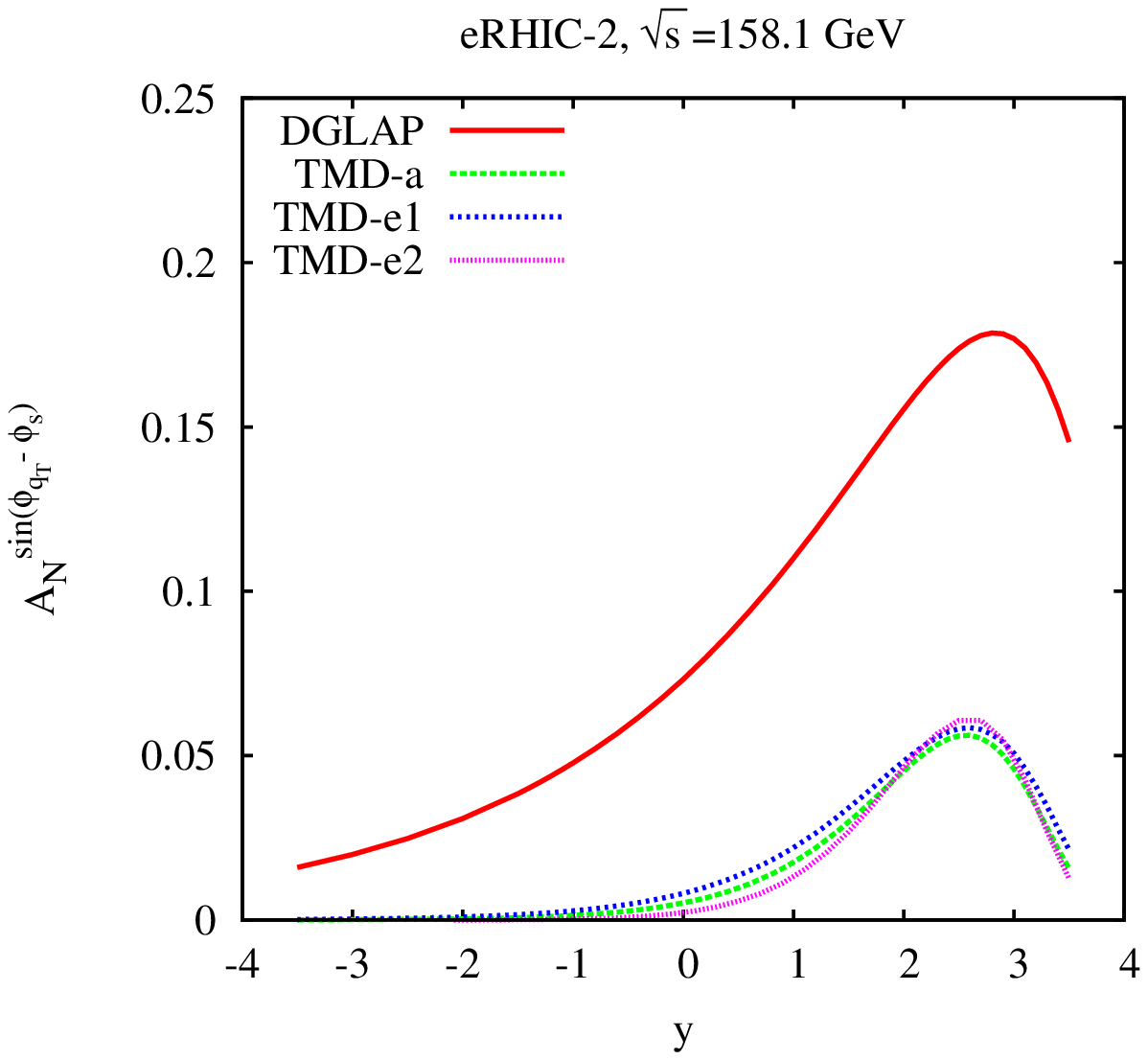}
\includegraphics[width=0.32\linewidth,angle=0]{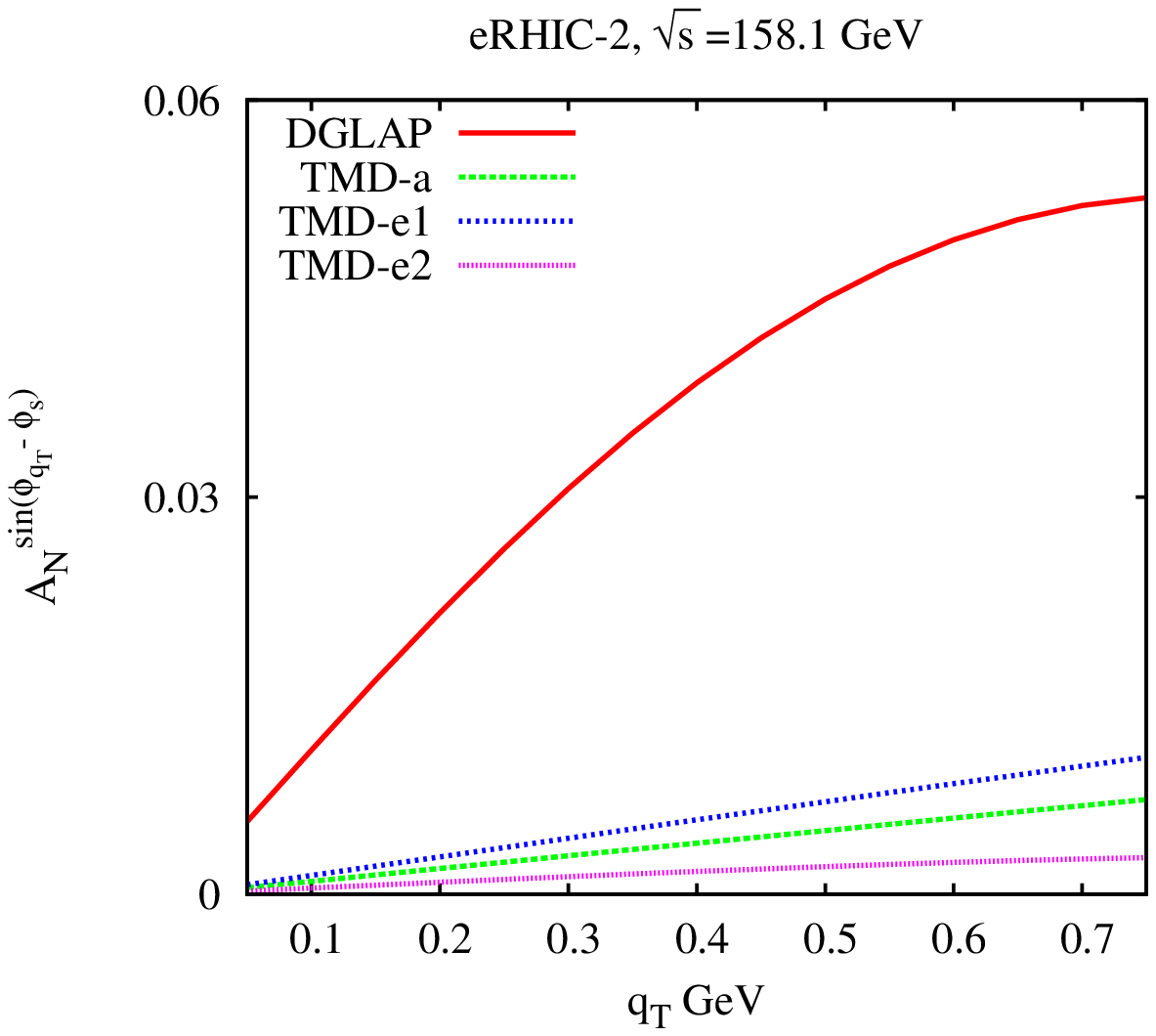}
\caption{eRHIC energy ($\sqrt{s} = 158.1$ GeV), Asymmetry as a function of $y$ (left panel) and $q_T$ (right panel). The integration ranges are $(0 \leq q_T \leq 1)$ GeV and $(-3.7 \leq y \leq 3.7)$\cite{Godbole:2014tha}. }
\label{erhic2_a}
\end{center}
\end{figure}
\begin{figure}[h]
\begin{center}
\includegraphics[width=0.32\linewidth,angle=0]{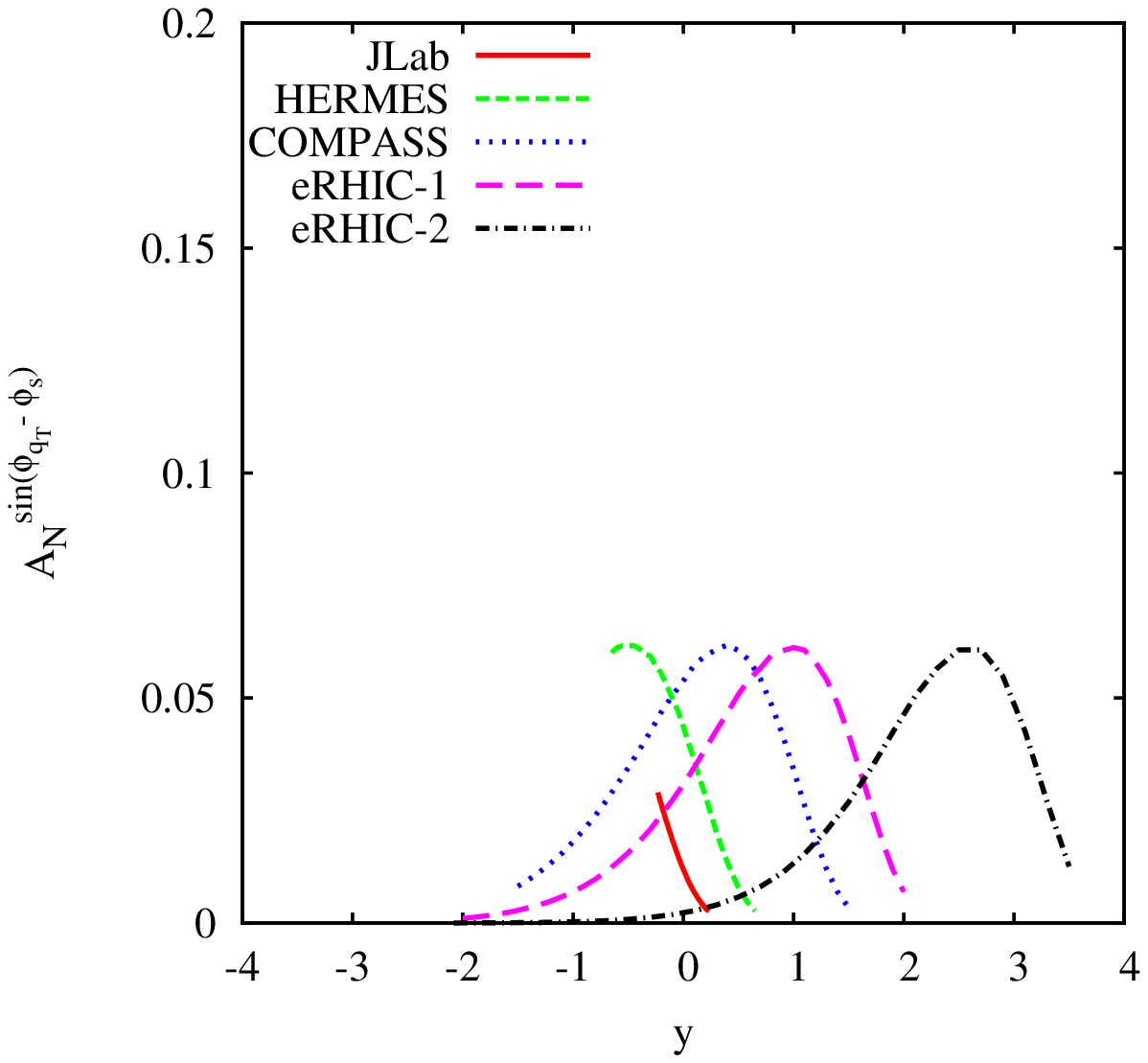}
\includegraphics[width=0.32\linewidth,angle=0]{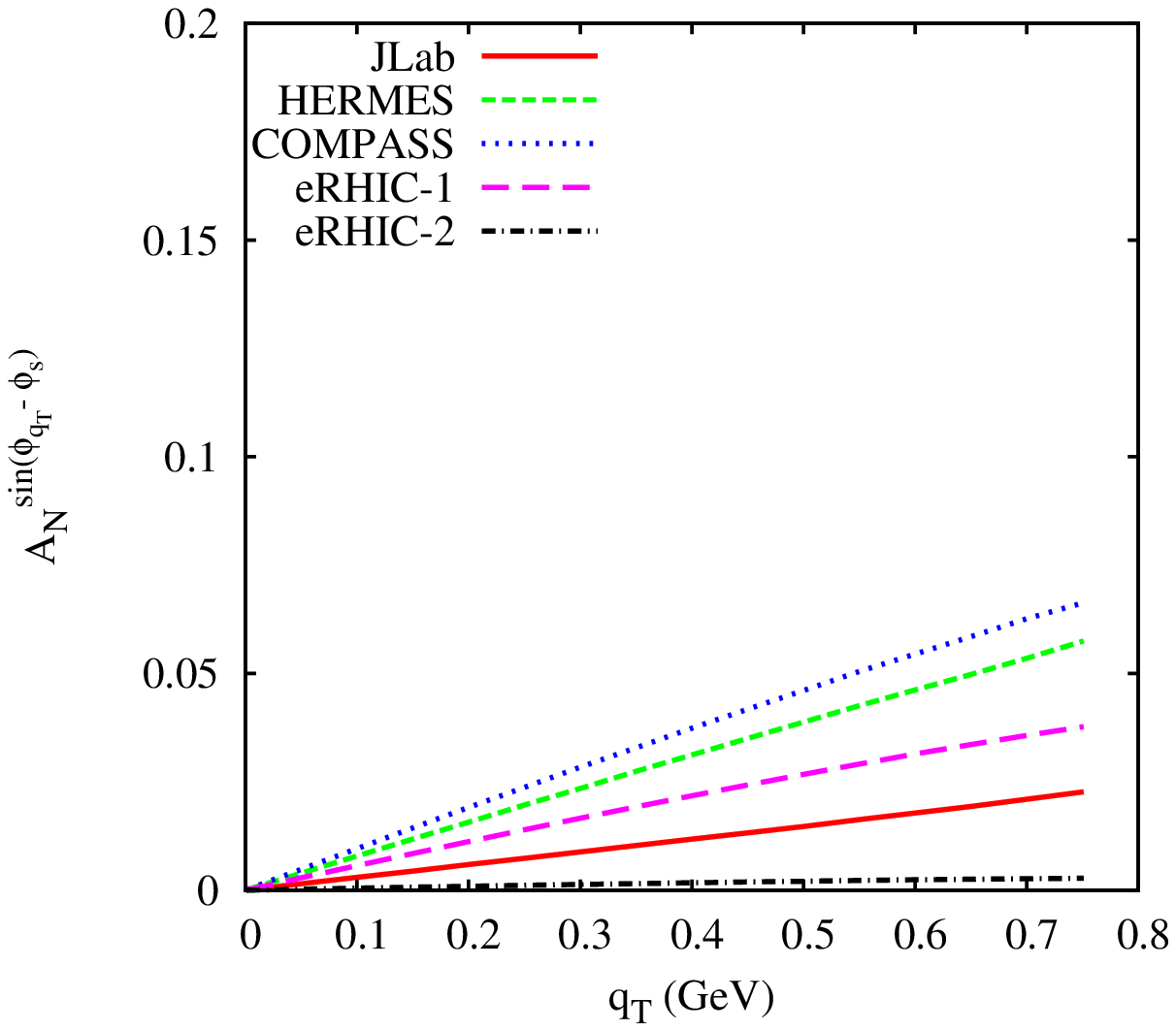}
\caption{Left panel: Plot of the Sivers asymmetry  in the $y$ distribution at  all c.o.m energies using the TMD-e2 fit .This plot shows the drift of the asymmetry peak towards higher values of rapidity $y$. Right panel: Plot of the Sivers Asymmetry  in the $q_T$ distribution\label{compare_y}}
\end{center}
\end{figure}

\FloatBarrier
\section{Numerical Estimates}
We will now present our estimates of SSA in photoproduction of $J/\psi$ for JLAB, HERMES, COMPASS and 
eRHIC energies. A detailed discussion of results can be found in Ref. \cite{Godbole:2014tha}.
Figs. 1-5 show the y and $k_T$ distribution for  different experiments  with parameterizations TMD Exact-1, TMD Exact -2 and TMD as given in Table \ref{parameter:set}. 
TMD-e1 parameter set, extracted at $Q_0 = 1.0\text{ GeV}$, is for the exact solution of TMD evolution equations extracted in Ref. \cite{Anselmino:2012aa}. TMD-a is the parameter set fitted to  analytical approximated solution of the Sivers function extracted in Ref.~\cite{Anselmino:2012aa}.
For estimates using NLL kernel, we have used the most recent parameters by Echevarria {\it et al.}\cite{Echevarria:2014xaa}
obtained by performing a global fit of all experimental data on Sivers asymmetry in SIDIS from HERMES, COMPASS 
and JLab. We call this set TMD-e2. This set was fitted at $Q_0 = \sqrt{2.4}\text{ GeV}$. 
Fig. 6 shows a comparison of asymmetries at all energies. 

 \section{Summary}
We have compared estimates of SSA in electroproduction of $J/\psi$ using  TMD's evolved via DGLAP evolution and TMD evolution schemes. 
For the latter, we have chosen three different parameter  sets fitted using  an approximate analytical solution, an exact solution at LL and an exact solution at NLL. 
We find that the estimates given by TMD evolved PDF's and Sivers function are all comparable but substantially small as compared to estimates calculated using DGLAP evolved TMD's.
\begin{acknowledgements}
A.M.  would like to thank the organizers of  LC2014 and Department of Physics, NCSU for their warm hospitality and 
University of Mumbai, University Grants Commission, India and Indian National Science Academy for financial support. This work was done 
under Grant No 2010/37P/47/BRNS of Department of Atomic Energy, India.  R.M.G. wishes to acknowledge support from the Department of Science and
Technology, India under Grant No. SR/S2/JCB-64/2007 under the J.C. Bose 
Fellowship scheme.
\end{acknowledgements}

%

\end{document}